# EQUATION OF STATE AND DISTRIBUTION OF PARTICLE SIZES IN GIBBS SYSTEM

V. V. Ryazanov

Institute for Nuclear Research, Kiev, Ukraine, e-mail: vryazan19@gmail.com

In the framework of Gibbs statistical theory, the issue of the distribution of particle sizes forming the statistical system and the moments of this distribution are considered. This task is relevant for a wide variety of applications. The distribution for particle sizes and moments of this quantity are determined from probabilistic considerations. Particle size depends on interactions in the system, on the compressibility factor, on the number of interacting particles, on the volume of the system. The expressions for the intrinsic volume of particles are substituted into the equations of state written using the theory of excluded volume for various expressions of the exclusion factor. The equations of state thus obtained can be considered as a refinement of the equation of state, a transition to a higher level of description.

# УРАВНЕНИЕ СОСТОЯНИЯ И РАСПРЕДЕЛЕНИЕ РАЗМЕРОВ ЧАСТИЦ В ГИББСОВСКОЙ СИСТЕМЕ

## В.В. Рязанов

*Институт ядерных исследований Национальной академии наук Украины,*

*03680, Киев, просп. Науки, 47, Украина*

E-mail: vryazan@yandex.ru

В рамках гиббсовской статистической теории рассмотрен вопрос распределения размеров частиц, образующих статистическую систему, и моментов этого распределения. Эта задача актуальна для самых разнообразных приложений. Распределение для размеров частиц и моменты этой величины определяются из вероятностных соображений. Размер частиц зависит от взаимодействий в системе, от фактора сжимаемости, от числа взаимодействующих частиц, от объема системы. Выражения для собственного объема частиц подставляются в уравнения состояния, записанные при помощи теории исключенного объема для различных выражений фактора исключения. Полученные таким образом уравнения состояния могут рассматриваться, как детализация уравнения состояния, переход на более высокий уровень описания.

*Ключевые слова:* большой канонический ансамбль, гиббсовская статистическая физика, распределение размеров частиц системы, уравнение состояния.

## ВВЕДЕНИЕ

Уравнение состояния и термодинамические свойства веществ важны в самых различных применениях. Известно большое число уравнений состояния, в основном эмпирических. Строгий вывод такого рода соотношений из статистической физики в научной литературе не так распространен. Уравнение состояния связывает между собой параметры состояния вещества (температура, давление, плотность, химический состав и пр.). Реальные уравнения состояний реальных веществ могут быть крайне сложными. Уравнение состояния системы не может быть выведено из термодинамики. Оно должно быть взято со стороны (из опыта или из модели, созданной в рамках статистической физики). По поводу моделей статистической физики в [1] сказано, что «Основная трудность последовательного теоретического расчета уравнения состояния вещества методами статистической физики заключается в необходимости корректного учета сложного по структуре межчастичного взаимодействия…». В настоящей работе эта задача также основная. Сделаны предположения о парности и аддитивности взаимодействия и использована концепция исключенного объема, конечности размеров частиц, оказавшаяся очень успешной в уравнении Ван-дер-Ваальса.

Существенным параметром статистической модели служит размер частиц (например, [2-4]). В многочисленных задачах спектроскопии, физической химии, в моделях конденсированного состояния и т.д. возникают вопросы, связанные с определением конечного размера частиц, образующих исследуемую систему. Так, проблема учета конечного объема частиц в связи с исследованием уравнения состояния ядерной материи обсуждалась в [5]. Эффекты исключенного объема (объема, недоступного для центров частиц из-за наличия собственного конечного объема частиц) оказываются важными в исследованиях многокомпонентного адронного газа [6]. В [7] получено уравнение состояния газов, представляющее собой свернутую форму вириального уравнения состояния, выраженную через экспоненциальные функции вириальных коэффициентов. Вириальные коэффициенты можно связать с размером частиц. В [8] показано, что некоторые термодинамические соотношения, вытекающие из уравнения Ван-дер-Ваальса (зависимость от размера частиц в котором существенна), могут быть применимы к реальным веществам и модельным системам, описываемым совершенно другими уравнениями состояния.

Связь теории исключенного объема с уравнениями состояния молекулярных систем обсуждается в [9], где концепция исключения объема или площади (для задач поверхности) называется главной. В [9] получен ряд уравнений состояния, предложен алгоритм их получения, основанный на понятиях исключенного объема, связанного с собственным объемом частиц, образующих систему. В основу первых уравнений состояния для флюида легли идеи Больцмана и Ван-дер-Ваальса об исключенном объеме.

Собственные объемы частиц играют важную роль в исследовании вязкости дисперсных систем [10]. В [11] отмечено, что небольшие изменения радиуса твердой сердцевины частиц или энергии парного взаимодействия могут приводить к заметным изменениям структуры и плотности жидкости. Для решения вопроса о причине высокой плотности гидроциклогексана следует обратиться к законам статистической физики.

Знание размеров частиц важно и во многих задачах биофизики (например [12]). Квазихимическая теория [13] также учитывает конечность объема частиц. Распределение занятого объема для системы твердых сфер получено в [14]. Можно найти сходство и отличие между результатами этой работы и настоящей работой.

Подходы к учету эффектов исключенного объема были предложены в работах [15-17]. Авторами ставилась цель моделирования поведения основных термодинамических величин при наличии у частиц собственных размеров. В работе [18] рассматривались эффекты конечного размера частиц в приближении среднего поля. В [19] рассматривалась задача о плотности вероятности распределения ближайших соседей. В [20] найдено геометрическое условие для предела плотности упаковки в неупорядоченных ансамблях одинаковых жестких выпуклых частиц. В [21] построены аналитические представления модуля всестороннего сжатия, уравнения состояния и удельного логарифма конфигурационного интеграла с учетом наличия у классических частиц эффективного размера (обусловленного отталкиванием на близких расстояниях).

В настоящей работе проблема собственных размеров частиц решается в общем виде для произвольных гиббсовских систем. Размер частиц связывается со статистическими и термодинамическими характеристиками системы. Во втором разделе получены выражения для распределения, первого и второго моментов размеров частиц. В третьем разделе полученные выражения для среднего значения размера частиц используются в качестве собственного размера частиц в уравнениях состояния.

## ФУНКЦИЯ РАСПРЕДЕЛЕНИЯ РАЗМЕРОВ ЧАСТИЦ

Плотность вероятности того, что в формализме большого канонического ансамбля Гиббса в конфигурационном пространстве статистической системы объемом $V$ содержится $n$ частиц с центрами тяжести в точках $r_1, ..., r_n$, равна (например, [2])

$$D_n(r_1,...,r_n; V; z) = \frac{z_1 \times ... \times z_n \exp\{-U_n/k_B T\}}{n! Q(z, V, T)}, \quad (1)$$

где $z_k \sim \exp\{\mu_k/k_B T\}$ - активность частицы в $k$-й точке, $\mu_k$ - химический потенциал этой частицы, $k_B$ - постоянная Больцмана, $T$ - абсолютная температура, $U_n$ - потенциальная энергия системы из $n$ частиц, $Q(z;V,T)$ - большая статистическая сумма. В отсутствии внешнего поля и в приближении парно-аддитивного взаимодействия

$$U_n = \sum_{i \neq j = 1}^{n} \varphi_{ij}/2; \quad \varphi_{ij} = \varphi(|r_i - r_j|); \quad y_i = \exp\{-\varphi_{ir}/k_B T\}; \quad y_{ij} = \exp\{-\varphi_{ij}/k_B T\}, \quad (2)$$

где $\varphi_{ij}$ - потенциал парного взаимодействия между частицами в точках $r_i$ и $r_j$. Присутствие внешнего поля $U_{ext}(r)$ можно учесть заменой химического потенциала $\mu_k$ на $\mu_k + U_{ext}(r_k)$ в активности частиц $z_k$. Большая статистическая сумма равна

$$Q(z;V,T) = 1 + \int_V z_1 dr_1 + \int_V \int_V z_1 z_2 y_{12} dr_1 dr_2 / 2! + ... = \exp\{p(z)V/k_B T\}, \quad (3)$$

где $p(z)$ - давление в системе с активностью $z$. Параметрами большого канонического ансамбля служат химический потенциал $\mu_k$ и потенциал взаимодействия $\varphi_{ij}$ или связанные с ними величины $z_k$ и $y_{ij}$; температура $T$, объем системы $V$.

Хотя во многих задачах статистической механики частицы считаются точечными, они обладают конечным объемом, который явно присутствует, например, в уравнении Ван-дер-Ваальса. В статистической механике это учитывается аналитическим видом функции потенциала взаимодействия $\varphi(r)$, ее стремлением к бесконечности при $r \to 0$, введением твердой сердцевины и т. д. [2-4]. Диаметром молекулы принято считать минимальное расстояние, на которое им позволяют сблизиться силы отталкивания. В настоящей работе приводятся вероятностные соображения, позволяющие связать параметры гиббсовского распределения с собственным или эффективным объемом составляющих систему частиц.

Известны выражения для функций распределения (корреляционных функций) гиббсовской системы [2], [22]

$$\rho^{(n)}(r_1,...,r_n;V;z) = \prod_{k=1}^{n} z_k e^{-\frac{U_n}{k_BT}} \frac{Q(z\prod_{k=1}^{n} y_k;V)}{Q(z,V)}; \quad (4)$$

$\rho^{(n)}(r_1,...,r_n;V;z)dr_1...dr_n$ является вероятностью того, что при наблюдении за системой, состоящей из $N$ молекул, мы найдем одну молекулу (не обязательно молекулу $1$) в элементе объема $dr_1$ около $r_1$, другую молекулу в элементе объема $dr_2$ около $r_2$,...и, наконец, последнюю молекулу в элементе объема $dr_n$ около $r_n$ [2]. Введем функции $\rho^{(n)}(r_1,...,r_n;\Delta_1,...,\Delta_n;V;z)$, представляющие собой совместную плотность вероятности того, что в системе находится $n$ частиц объемом $\Delta_1,...,\Delta_n$ с центрами в точках $r_1,...,r_n$. Предположив независимость событий расположения центров частиц в точках $r_1,...,r_n$ и того, что частица с центром в точке $r_k$ имеет объем $\Delta_k$, запишем соотношение

$$\rho^{(n)}(r_1,...,r_n;\Delta_1,...,\Delta_n;V;z) \sim \rho^{(n)}(r_1,...,r_n;V;z)P^{(n)}(\Delta_1,...,\Delta_n), \quad (5)$$

где $P^{(n)}(\Delta_1,...,\Delta_n)$ - функция распределения размеров частиц в связанной группе из $n$ частиц.

Событие, состоящее в том, что в системе объемом $V$ находится частица, можно представить, как пересечение событий; происходят два события: первое - в системе объемом $V$ находится частица с эффективным объемом $\Delta(r_1)$ с центром тяжести в точке $r_1$, второе - в системе объемом $V-\Delta(r_1)$ нет частиц. Одну и ту же частицу (или группу частиц) можно считать как принадлежащей исследуемой системе, так и внешней по отношению к системе без этой группы частиц. Эти два события можно считать независимыми. Но в системе объемом $V-\Delta(r_1)$ присутствует внешнее поле системы объема $\Delta(r_1)$, которое создает находящаяся там частица. Такие рассуждения приводят к соотношениям

$$D_1(r_1;V;z) = \rho^{(1)}(r_1;\Delta_1)D_0(V-\Delta_1;zy_1); \quad (6)$$

$$D_2(r_1,r_2;V;z) = 2^{-1}\rho^{(1)}(r_1;\Delta_1)D_1(r_2;V-\Delta_1;zy_1) = 2^{-1}\rho^{(2)}(r_1,r_2;\Delta_1,\Delta_2)D_0(V-\Delta_1-\Delta_2;zy_1y_2).$$

Таким же образом для группы (кластера) из $n$ взаимодействующих частиц можно связать вероятность того, что в системе объемом $V$ находится $n$ частиц, и вероятность того, что в системе объемом $V-\Delta_1-...-\Delta_m$ находится $n-m$ ($n \geq m$) частиц, где $\Delta_k$ - эффективный объем одной $k$-й частицы, соотношениями

$$D_n(r_1,...,r_n;V;z) = n^{-1}\rho^{(1)}(r_1;\Delta_1)D_{n-1}(r_2,...,r_n;V-\Delta_1;zy_1) = [n(n-1)]^{-1}\rho^{(2)}(r_1,r_2;\Delta_1,\Delta_2) \times$$

$$D_{n-2}(r_3,...,r_n;V-\Delta_1-\Delta_2;zy_1y_2) = ... = (n!)^{-1}\rho^{(n)}(r_1,...r_n;\Delta_1,...,\Delta_n)D_0(V-\sum_{k=1}^{n}\Delta_k; z\prod_{k=1}^{n}y_k). \quad (7)$$

Влияние частиц, исключающихся из системы, учитывается как внешнее поле, добавки к химическому потенциалу. Это не приближение «самосогласованного поля», а точный учет взаимодействий в выбранных подсистемах. Множители $(n)^{-1}$, $[n(n-1)]^{-1},...,(n!)^{-1}$ учитывают неразличимость частиц, то, что удаляться может любая из $n, n-1,...$ частиц (в $D_{n-1}$, например, может отсутствовать любая из $n$ частиц, не обязательно $n$-я).

Подставляя выражения (1) в соотношения (6), (7) получим (при этом учтем вытекающее из (1) выражение: $D_0(V;z,T)=1/Q(z;V,T)$):

$$\rho^{(n)}(r_1,...,r_n;\Delta_1,...,\Delta_n;V;z) = \prod_{k=1}^{n} z_k e^{-\frac{U_n}{k_B T}} \frac{Q(z\prod_{k=1}^{n} y_k; V-\sum_{k=1}^{n}\Delta_k)}{Q(z;V)} \quad . \quad (8)$$

Подставляя в выражение (8) соотношения (5) и (4), получаем, что функция распределения размеров частиц в связанной группе из $n$ частиц $P^{(n)}(\Delta_1,...,\Delta_n)$ пропорциональна статистической сумме с измененными активностями и объемом

$$P^{(n)}(\Delta_1,...,\Delta_n) \sim Q(z\prod_{k=1}^{n} y_k; V-\sum_{k=1}^{n}\Delta_k) = \exp\{\frac{p(z\prod_{k=1}^{n} y_k)[V-\sum_{k=1}^{n}\Delta_k]}{k_B T}\}, \quad (9)$$

где $p(Az) = k_B T \ln Q(Az;V;T)/V$ - давление при активности $Az$ (см. выражение (3)).

Плотность вероятности функции распределения $P^{(n)}(\Delta_1,...,\Delta_n)$ определяется из (9)

$$p^{(n)}(\Delta_1,...,\Delta_n) = \frac{1}{B^{(n)}} \frac{\partial^n Q(z\prod_{k=1}^{n} y_k; V-\sum_{k=1}^{n}\Delta_k)}{\partial\Delta_1...\partial\Delta_n}, \quad B^{(n)} = \int...\int \frac{\partial^n Q(z\prod_{k=1}^{n} y_k; V-\sum_{k=1}^{n}\Delta_k)}{\partial\Delta_1...\partial\Delta_n} d\Delta_1...d\Delta_n, \quad (10)$$

где $B^{(n)}$ - нормировочный множитель. Интегрирование в $B^{(n)}$ должно проводиться с учетом того, что величины $\Delta_1,...,\Delta_n$ занимают определенный объем, т.е. по правилу

$$\int...\int(...)d\Delta_1...d\Delta_n = \int_0^V d\Delta_1 \int_0^{V-\Delta_1} d\Delta_2 ... \int_0^{V-\Delta_1-...-\Delta_{n-1}}(...)d\Delta_n. \quad (11)$$

При этом

$$\frac{\partial^n Q(z\prod_{k=1}^{n} y_k; V-\sum_{k=1}^{n}\Delta_k)}{\partial\Delta_1...\partial\Delta_n} = (-\alpha_n)^n e^{\alpha_n(V-\Delta_1-...-\Delta_n)}; \quad \alpha_n = p(zy_1...y_n)/k_B T; \quad B^{(n)} = (-1)^{n-1}[\sum_{k=0}^{n-1}\frac{V^k \alpha_n^k}{k!} - \exp\{\alpha_n V\}].$$

Из (9)-(10) получаем, что плотность вероятности $p^{(n)}(\Delta_1,...,\Delta_n)$ равна

$$p^{(n)}(\Delta_1,...,\Delta_n) = \frac{(-\alpha_n)^n \exp\{\alpha_n(V-\Delta_1-...-\Delta_n)\}}{B^{(n)}} \quad .$$

Интегрирование вида (11) и соответствующая нормировка отличают распределение (12) от показательного распределения. При $n=1, 2, 3...$:

$$p^{(1)}(\Delta_1) = \frac{\alpha_1 \exp\{-\alpha_1 \Delta_1\}}{1-\exp\{-\alpha_1 V\}}; \qquad p^{(2)}(\Delta_1,\Delta_2) = \frac{\alpha_2^2 \exp\{-\alpha_2(\Delta_1+\Delta_2)\}}{1-\exp\{-\alpha_2 V\}(1+\alpha_2 V)};\ldots;$$

$$p^{(r)}(\Delta_1,\ldots,\Delta_r) = \frac{\alpha_r^r \exp\{-\alpha_r(\Delta_1+\Delta_2+\ldots+\Delta_r)\}}{1-\exp\{-\alpha_r V\}\sum_{k=0}^{r-1}\frac{(\alpha_r V)^k}{k!}}; \qquad r=1,2,\ldots. \qquad (12)$$

Используя формулу для разложения Тейлора [23]:

$$e^{\alpha_r V} = \sum_{k=0}^{r} \frac{(\alpha_r V)^k}{k!} + R_r; \qquad R_r(\alpha_r V) = \frac{(\alpha_r V)^{r+1}}{(r+1)!} e^{\theta_r \alpha_r V}; \qquad 0 < \theta_r < 1,$$

где $R_r(\alpha_r V)$ - остаточный член ряда Маклорена (форма Лагранжа остаточного члена), получаем: $B^{(n)} = (-1)^n R_{n-1}(\alpha_n V)$; $p^{(r)}(\Delta_1,\ldots,\Delta_r) = \frac{\alpha_r^r}{R_{r-1}(\alpha_r V)} e^{\alpha_r(V-\Delta_1-\ldots-\Delta_r)} = \frac{r!}{V^r} e^{\alpha_r[V(1-\theta_{r-1})-\Delta_1-\ldots-\Delta_r]}$.

В [23] остаточный член, когда $n$ – целое число, записан в виде

$$R_n(x) = \int_0^x \frac{e^t}{n!}(x-t)^n dt = \gamma(n+1,x)\frac{e^x}{n!}, \quad \gamma(n+1,x) = n!(1-e^{-x}\sum_{k=0}^{n}\frac{x^k}{k!}), \quad R_n(x) = e^x(1-e^{-x}\sum_{k=0}^{n}\frac{x^k}{k!}), \quad (13)$$

где $\gamma(\alpha,x)$ - неполная гамма-функция [23]. Из (12)-(13) получаем, что

$$p^{(r)}(\Delta_1,\ldots,\Delta_r) = \frac{\alpha_r^r (r-1)!}{\gamma(r,\alpha_r V)} \exp\{-\alpha_r(\Delta_1+\ldots+\Delta_r)\}. \qquad (14)$$

Для целых значений $n$ получаем из (13)-(14) выражение (12). Выражение (14) может описывать комплексы частиц и в случае дробных значений их числа, величины $r$.

Если определить при помощи выражений (10)-(14) среднее значение объема частицы $\Delta_1$ в группе из $n$ связанных частиц

$$\overline{\Delta_1^{(n)}} = \int\ldots\int \Delta_1 p^{(n)}(\Delta_1,\ldots,\Delta_n) d\Delta_1\ldots d\Delta_n,$$

то получим (интегрирование проводится по правилу (11)) величину среднего значения

$$\overline{\Delta_1^{(n)}} = \frac{1}{\alpha_n}\frac{R_n(\alpha_n V)}{R_{n-1}(\alpha_n V)} = \frac{1}{\alpha_n}\frac{1-\exp\{-\alpha_n V\}\sum_{k=0}^{n}\frac{(\alpha_n V)^k}{k!}}{1-\exp\{-\alpha_n V\}\sum_{k=0}^{n-1}\frac{(\alpha_n V)^k}{k!}} = \frac{1}{\alpha_n}[1 - \frac{\exp\{-\alpha_n V\}\frac{(\alpha_n V)^n}{n!}}{1-\exp\{-\alpha_n V\}\sum_{k=0}^{n-1}\frac{(\alpha_n V)^k}{k!}}], \quad (15)$$

$\overline{\Delta_1^{(n)}} = k_n/(\alpha_0 + c_n)$, $\alpha_0 = p(z)/k_B T$, $\alpha_n = \alpha_0 + c_n$, $k_n = R_n(\alpha_n V)/R_{n-1}(\alpha_n V)$. При $n=1$

$$\overline{\Delta_1^{(1)}} = \frac{[1-(1+\alpha_1 V)\exp\{-\alpha_1 V\}]}{\alpha_1(1-\exp\{-\alpha_1 V\})}; \qquad \overline{\Delta_1^{(n)}} = \frac{V}{(n+1)} e^{-\frac{p(zy_1\ldots y_n)V(\theta_{n-1}-\theta_n)}{k_B T}} = \frac{V}{(n+1)} e^{-\alpha_n V(\theta_{n-1}-\theta_n)}. \quad (16)$$

При объеме системы $V \to \infty$, $\overline{\Delta_1^{(1)}} = 1/\alpha_1$; для идеального газа $\overline{\Delta_1^{(1)}} = v = V/\overline{N} = 1/\rho$, где $\overline{N}$ - полное среднее число частиц в системе, $\rho$ - средняя плотность; при $V \to 0$, $\overline{\Delta_1^{(1)}} \to 0$, что согласуется с физическими представлениями. Так же и $\overline{\Delta_1^{(n)}} = 1/\alpha_n$ при $V \to \infty$; для идеального газа $\overline{\Delta_1^{(n)}} = v = V/\overline{N} = 1/\rho$. При $V \to 0$, $\overline{\Delta_1^{(n)}} \to 0$. При $n \to \infty$, $\overline{\Delta_1^{(n)}} \to 0$. Если $n \to \infty$ и $V \to \infty$, то $\overline{\Delta_1^{(n)}} \to V/n$. При $\alpha_n \to 0$, $\overline{\Delta_1^{(n)}} \to V/n$; $\alpha_n \to \infty$, $\overline{\Delta_1^{(n)}} \to 0$.

Дифференцируя выражение (15) по $\alpha_n$, получим, что

$$\frac{\partial \alpha_n}{\partial V} = \frac{\alpha_n}{t_R^{-1}(\overline{\Delta_1^{(n)}} + \alpha_n \frac{\partial \overline{\Delta_1^{(n)}}}{\partial \alpha_n}) - V}; \quad t_R = \frac{\partial (R_n(x)/R_{n-1}(x))}{\partial x} = \frac{[R^2_{n-1}(x) - R_n(x)R_{n-2}(x)]}{R^2_{n-1}(x)}\bigg|_{x=\alpha_n V}.$$

Фазовый переход происходит при $\frac{\partial p}{\partial v} \geq 0$; $v = \frac{V}{\overline{N}}$. Если среднее полное число частиц в системе $\overline{N}$ постоянно, то $\frac{\partial \alpha_n}{\partial V} \sim \frac{\partial p}{\partial v}$. Величина $\alpha_n > 0$, и фазовый переход наступает при

$$t_R^{-1}(\overline{\Delta_1^{(n)}} + \alpha_n \frac{\partial \overline{\Delta_1^{(n)}}}{\partial \alpha_n}) - V \geq 0.$$

Явная зависимость $\theta_n$ и $\theta_{n-1}$ от термодинамических параметров не определяется из вида остаточного члена. Поэтому выражение остаточного члена через неполную гамма-функцию может быть предпочтительней.

Из соотношений (10)-(14) определяется также дисперсия значения $\Delta_1^{(n)}$ в виде

$$D^{(n)}_\Delta = \overline{\Delta^{(n)2}_1} - (\overline{\Delta^{(n)}_1})^2 = \frac{1}{\alpha_n^2}[\frac{2R_{n+1}(\alpha_n V)}{R_{n-1}(\alpha_n V)} - (\frac{R_n(\alpha_n V)}{R_{n-1}(\alpha_n V)})^2]. \qquad (17)$$

При $n=1$: $D_\Delta^{(1)} = \overline{\Delta^{(1)2}_1} - (\overline{\Delta^{(1)}_1})^2 = \frac{e^{2\alpha_1 V} + 1 - (\alpha_1 V)^2 e^{\alpha_1 V} - 2e^{\alpha_1 V}}{\alpha_1^2(e^{2\alpha_1 V} + 1 - 2e^{\alpha_1 V})} \approx \frac{1}{\alpha_1^2}$, при больших $V$. Для величины $\overline{\Delta^{(n)2}_1}$ получаем

$$\overline{\Delta^{(n)2}_1} = \frac{2}{\alpha_n^2}\frac{R_{n+1}[\alpha_n V]}{R_{n-1}[\alpha_n V]} = \frac{2}{\alpha_n^2}(1 - \frac{e^{-x}\frac{x^n}{n!}(1+\frac{x}{n+1})}{1 - e^{-x}(1+x+...+\frac{x^{n+1}}{(n+1)!})})\bigg|_{x=\alpha_n V} = \frac{2V^2}{(n+2)(n+1)}e^{-\alpha_n V(\theta_{n-1}-\theta_{n+1})}.$$

Из (10)-(17) записывается соотношение $\overline{\Delta^{(n)}_1 + ... + \Delta^{(n)}_n} = \overline{\Delta^{(n)}_1} + ... + \overline{\Delta^{(n)}_n}$, корреляторы вида $\langle \Delta^{(n)}_1 \Delta^{(n)}_2 \rangle$, а также другие соотношения и функции от объемов частиц $\Delta_1, ..., \Delta_n$.

Величину $\alpha_n = p(zy_1...y_n)/k_B T$ можно связать с корреляционными функциями $\rho^{(n)}$. Из (3), (4) находим (соотношения, связывающие $\alpha_n$ с $\alpha_0$, получены также в [2])

$$\rho^{(n)}(r_1,...,r_n;V;z) = \prod_{k=1}^n z_k e^{-U_n/k_B T + (\alpha_n - \alpha_0)V}; \quad \alpha_n = \alpha_0 + [\ln(\rho^{(n)}/z_1...z_n) + U_n/k_B T]/V; \quad \alpha_0 = p(z)/k_B T; \quad \alpha_n = \alpha_0 + c_n. \quad (18)$$

Можно записать приближенные выражения для величин $c_n$ и $\alpha_n$. Интеграл от $\rho^{(n)}$ для больших значений объема $V$ в соответствии с [2] равен

$$\lim_{V \to \infty} \int_V ... \int \rho^{(n)}(r_1,...,r_n) dr_1...dr_n = V^n \rho^n. \qquad (19)$$

Используем полученное в [2] выражение

$$e^{(p-p^*)V/k_BT} = \sum_{m\geq 0} \frac{(z-z^*)^m}{m!} \left(\frac{1}{z^*}\right)^m \int_V \ldots \int \rho^{(m)}(r_1,\ldots,r_m,z^*) dr_1 \ldots dr_m, \qquad (20)$$

где $p^* = p(z^*), \rho^* = \rho(z^*)$. Положим $z^* = z y_{f1} y_{f2} \ldots y_{fn}$, как в $\alpha_n$, где $f$ - некоторая точка в объеме системы. Если в (20) подставить выражение (19), то получаем соотношение

$$\frac{(p-p^*)}{k_BT} = \frac{(z-z^*)}{z^*}\rho^*; \quad \frac{p(z^*)}{k_BT} = \frac{p(z)}{k_BT} + \left(1 - \frac{1}{y_{f1}\ldots y_{fn}}\right)\rho(z^*). \qquad (21)$$

Еще одно, полученное в [2], выражение имеет вид

$$e^{(p-p^*)V/k_BT} \frac{z^*}{z} \rho^{(1)}(r_1,z) = \sum_{m\geq 0} \frac{(z-z^*)^m}{m!} \left(\frac{1}{z^*}\right)^m \int_V \ldots \int \rho^{(m+1)}(r_1,\ldots,r_{m+1},z^*) dr_2 \ldots dr_{m+1}. \qquad (22)$$

Интегрируя это соотношение по $r_1$, используя приближение (19) и сопоставляя с (20) получим, что

$$\rho^*(z^*) = \rho(z) z^*/z = y_{1f}\ldots y_{nf}\rho(z), \quad \rho(z) = \rho^{(1)}(r_f, z). \qquad (23)$$

Выражение (23) можно получить без использования приближения (19) в (20), (22), но в приближении $\rho^{(m+1)} = \rho^{(m)}\rho^{(1)}$. Подстановка выражения (23) в (21) дает

$$\frac{p(z^*)}{k_BT} = \frac{p(z)}{k_BT} + (y_{f1}\ldots y_{fn} - 1)\rho(z), \quad c_n = (y_{f1}\ldots y_{fn} - 1)\rho(z) = c_y\rho, \quad c_y = (y_{f1}\ldots y_{fn} - 1). \qquad (24)$$

Более подробно эти вопросы изложены в [24]. В [24] получены также выражения для моментов размеров частиц с жесткой несжимаемой сердцевиной объемом $d$ (т.е. $\Delta \geq d$; выше полагалось $d=0$). При произвольных $n$:

$$\overline{\Delta^{(n)}}_{1\,d} = \frac{1}{\alpha_n}\left(\frac{R_n[\alpha_n(V-nd)]}{R_{n-1}[\alpha_n(V-nd)]} + d\alpha_n\right) = d + \frac{1}{\alpha_n} \frac{1 - \exp\{-\alpha_n(V-nd)\}\sum_{k=0}^{n}\frac{(\alpha_n(V-nd))^k}{k!}}{1 - \exp\{-\alpha_n(V-nd)\}\sum_{k=0}^{n-1}\frac{(\alpha_n(V-nd))^k}{k!}}. \qquad (25)$$

При $d\to 0$ соотношения (25) переходят в (15). В выражениях для частиц с твердой сердцевиной явно присутствует размер частиц $d$ в отличие от выражений вида (15). Существенна зависимость от объема системы $V$ и числа взаимодействующих частиц $n$. Можно связать величину $\Delta$ с теорией свободного объема [2] и с решеточными теориями.

Величина $\overline{\Delta^{(n)}}_{1\,d}$ зависит от $n$ - числа частиц в группе взаимодействующих частиц. При этом $\overline{\Delta^{(n)}}_{1\,d} - d \leq 1/\alpha_n$; $\overline{\Delta^{(n)}}_{1\,d} - d = k_{nd}/\alpha_n$; $0 \leq k_{nd} = R_n[\alpha_n(V-nd)/R_{n-1}[\alpha_n(V-nd)] \leq 1$. При $\alpha_n \to 0$, $\overline{\Delta_1^{(n)}}_d \to V/n$, не зависит от $d$; $\alpha_n \to \infty$, $\overline{\Delta_1^{(n)}}_d \to d$. Максимальное значение $n^{max} = V/d$, т.к. $dn \leq V$. При $n \to n^{max}$, $\alpha_n V(1-nd/V) \to 0$; $\overline{\Delta^{(n)}}_{1\,d} \to d$. Величина $\overline{\Delta^{(n)}}_{1\,d} - d$ в зависимости от $n$ ведет себя, как показано на рис.1а-1б, где обозначено $r = \overline{\Delta^{(n)}}_{1\,d} - d$, $a = \alpha_n$. При расчете использовалась модель упругих шаров [3] с потенциалом вида $\Phi(r)=0$, $r>r_0$; $\Phi(r)=\infty$, $r \leq r_0 \sim d^{1/3}$. Тогда можно не учитывать зависимость $\alpha_n$ от $n$. При этом $\partial \overline{\Delta^{(n)}}_{1\,d}/\partial n \leq 0$. На рис.1б показана ситуация, когда при некоторых $n$ $\partial \overline{\Delta^{(n)}}_{1\,d}/\partial n > 0$. Для разреженных газов $n \sim 1$, значения $n$ велики для твердого тела, и $n$ занимает промежуточные значения для жидкостей. Можно найти значения $n$, которые соответствовали бы приближениям идеального газа, Ван-дер-Ваальса, аппроксимантам Паде, модели твердых сфер и т.д.

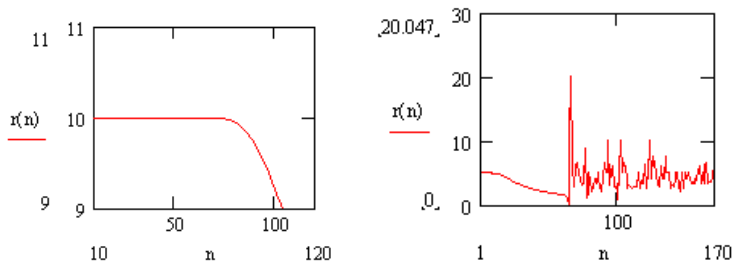

Рис.1а. n=10,…,120; a=$10^{-1}$; d=$10^{-5}$; V=$10^3$.　　Рис.1б. n=1,2,…,170; a=$10^{-0.7}$; d=$10^{-3}$; V=$10^2$; $\partial \overline{\Delta^{(n)}}_{1\,d}/\partial n > 0$, область фазового перехода.

Рис.1. Зависимость разности между средним размером частицы $\overline{\Delta^{(n)}}_{1\,d}$ в кластере $n$ взаимодействующих частиц в модели с жесткой несжимаемой сердцевиной объемом $d$ и величиной $d$ от $n$ - числа частиц в группе взаимодействующих частиц. Использована модель взаимодействия упругих шаров [3]; ($\partial \overline{\Delta^{(n)}}_{1\,d}/\partial n) \leq 0$; r= $\overline{\Delta^{(n)}}_{1\,d}$-d; a=$\alpha_n$=P(z$y_1$...$y_n$)/$k_B$T.

Рис. 1а: ($\partial \overline{\Delta^{(n)}}_{1\,d}/\partial n) \leq 0$; Рис. 1б: при некоторых n, ($\partial \overline{\Delta^{(n)}}_{1\,d}/\partial n) > 0$, фазовый переход, система находится между бинодалью и спинодалью.

Рассмотрим зависимость $\overline{\Delta^{(n)}}_{1\,d}$-$d$ от $\alpha_n$: при $\alpha_n \to 0$, $\overline{\Delta^{(n)}}_{1\,d} \to V/n$; при $\alpha_n \to \infty$, $\overline{\Delta^{(n)}}_{1\,d}$-$d \to 0$. Зависимость $\overline{\Delta^{(n)}}_{1\,d}$-$d$ от $\alpha_n$ ведет себя, как показано на рис.2. Так как $\alpha_n \to 0$ при $\varphi_{12} \to \infty$, а $\alpha_n \to \alpha_{n=0} = \alpha_0$ при $\varphi_{12} \to 0$, то так же (но $\overline{\Delta^{(n)}}_{1\,d}$-$d$ растет с ростом $\varphi_{12}$) ведет себя зависимость $\overline{\Delta^{(n)}}_{1\,d}$-$d$ от $y_{12}$. Возможны ситуации, при которых величина $\overline{\Delta^{(n)}}_{1\,d}$-$d$ может расти с ростом $\alpha_n$. Давление ($\sim \alpha_n \sim P/k_B T$) "сжимает" частицу до минимального значения, равного $d$.

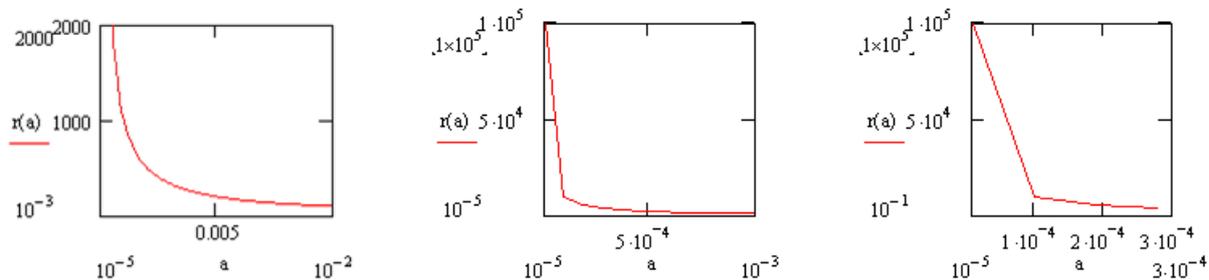

Рис.2а. a=$10^{-5}$,…,$10^{-2}$; d=$10^{-7}$; n=10; V=$10^{17}$.　Рис.2б. a=$10^{-5}$,…,$10^{-3}$; d=$10^{-5}$; V=$10^{10}$; n=30.　Рис.2в. a=$10^{-5}$,…,3·$10^{-4}$; d=$10^{-1}$; V=$10^{10}$; n=47.

Рис.2. Зависимость разности между средним размером частицы $\overline{\Delta^{(n)}}_{1\,d}$ в кластере $n$ взаимодействующих частиц в модели с жесткой несжимаемой сердцевиной объемом $d$ и величиной $d$ от $a=\alpha_n$ – фактора сжимаемости (умноженного на $R/k_B V$, $R$ – универсальная газовая постоянная); r= $\overline{\Delta^{(n)}}_{1\,d}$-d; a=$\alpha_n$=P(z$y_1$...$y_n$)/$k_B$T.

Ниже рассмотрены случаи макроскопических систем большого объема (для малых объемов уравнения состояния более сложные) и близкодействия, короткодействующего потенциала взаимодействия, когда значения *n* порядка 6, и не равны полному числу частиц в системе *N*. В этом случае выражения для $k_n$ (15), равные $R_n(\alpha_n V)/R_{n-1}(\alpha_n V)$, приблизительно равны 1. Даже если величины *n* порядка $10^2$-$10^3$, то все равно эти значения много меньше значений полного числа частиц *N* в макроскопических системах.

Оценим величину $k_n$ из (15).

$$k_n(x) = \frac{R_n(x)}{R_{n-1}(x)} = \frac{1 - \exp\{-x\} \sum_{k=0}^{n} \frac{(x)^k}{k!}}{1 - \exp\{-x\} \sum_{k=0}^{n-1} \frac{(x)^k}{k!}} = \frac{1}{1 - \Sigma_1^{-1}}, \quad x = \alpha_n V, \alpha_n(V-nd);$$

$\Sigma_1 = \frac{x}{n+1} + \frac{x^2}{(n+2)(n+1)} + ... + \frac{x^m}{(n+m)...(n+1)} + ...$ Используем соотношения [24]

$$\sum_{k=0}^{n} \frac{(x)^k}{k!} = \frac{\Gamma(n+1,x)e^x}{n!}, \quad \sum_{k=0}^{\infty} \frac{(x)^k}{(rk+m)!} = \frac{1}{rx^m}[\sum_{k=1}^{n} \frac{e^{\theta_k x^{1/n}}}{\theta_k^m} - r\sum_{k=1}^{[m/n]} \frac{(x)^{(m-rk)/r}}{(m-rk)!}], \quad (26)$$

$\theta_k$ один из корней степени *r* из 1: $\theta_k = \sqrt[r]{1} = \cos(2k\pi/r) + i\sin(2k\pi)r$. Положив в (26) *r=1*, *m=n+1*, получаем $\Sigma_1 = \frac{e^x n!}{x^n}[1 - \frac{\Gamma(n+2,x)}{(n+1)!}]$. При больших макроскопических объемах *V* $x \sim V$, $n \ll N$, $\Gamma(n+2,x) \to 0$, $\Sigma_1 \to \infty$, $k_n(x) \cong 1 - \Sigma_1^{-1} \approx 1$.

## УРАВНЕНИЕ СОСТОЯНИЯ И РАЗМЕР ЧАСТИЦ

В [9] использование концепции фактора исключения называется главным аспектом уравнения состояния. В [9] записывается производящее уравнение, генерирующее уравнения состояния. Решение этого уравнения зависит от фактора исключения (приведенного исключенного объема). Однако в полученные уравнения состояния, например, уравнение Ван-дер-Ваальса, входит собственный объем частиц, параметр, как правило, не очень точно определенный. В настоящей работе в качестве этого параметра рассматривается среднее значение случайной величины размера частиц, полученное в предыдущем разделе. Среднее значение величины размера частиц выражено через термодинамические параметры системы, и его подстановка в уравнения, записанные в [9], приводит к новым уравнениям. Эту подстановку явного вида выражения для среднего размера частиц можно рассматривать, как детализацию уравнения состояния, более высокий уровень описания. Знание выражений для размера частиц системы позволяет осуществить новый подход к уравнениям состояния.

Для простоты рассмотрим однокомпонентный случай. Можно провести обобщение на многокомпонентную систему. В [9] фактор исключения $f = v^{ex}/v_0$, отношение среднего исключенного объема $v^{ex}$ к среднему значению собственного объема частиц в системе $v_0$, записывается в виде функции от общей объемной доли всех компонентов $\varphi = \rho v_0$, $\rho$ - плотность числа частиц, $v_0$ - среднее значение собственного объема частиц. Будем в качестве этой величины рассматривать полученные средние значения размеров частиц из (15) или (25). При задании различных значений фактора исключения $f$, в [9] получены различные уравнения состояния. В [9] изменение фактора исключения связывается с кластеризацией вещества и с изменением упаковки частиц при повышении плотности. Но изменение фактора исключения можно рассматривать также как учет случайности в собственных размерах частиц, составляющих систему. И в [9] при помощи задания различного вида фактора исключения проводится дискретное задание случайного размера частиц. Эти изменения размеров частиц вызваны теми же причинами, но в предыдущем разделе они учитываются более строго, при помощи общих соотношений гиббсовской статистической механики. Производящее уравнение записано в [9] в виде

$$\tilde{p} = \frac{pv_0}{k_B T} = \int_0^{\varphi} \frac{d\varphi}{1 - f\varphi} - \frac{\alpha \rho^2 v_0}{k_B T}, \qquad (27)$$

где $\varphi = \rho v_0$, $\rho$ - плотность числа частиц $\alpha$ - постоянная сил притяжения частиц для сил Ван-дер-Ваальса, $v_0$ - собственный объем частиц.

Наиболее грубым (нулевым) приближением для производящего уравнения (27) в [9] служит предположение о постоянстве фактора исключения, когда для неионизированного однокомпонентного одноатомного газа $f = f_0 = 8$. Тогда уравнение состояния принимает вид уравнения Планка [9]

$$\tilde{p} = \frac{pv_0}{k_B T} = -\frac{1}{8}\ln(1 - 8\rho v_0) - \frac{\alpha \rho^2 v_0}{k_B T}. \qquad (28)$$

Приравняем величину собственного объема частиц $v_0$ к полученному в (15) среднему размеру частиц, и в соответствии с выражениями (15), (24), запишем

$$v_0 = \frac{k_n}{\alpha_0 + c_n}, \quad \alpha_0 = \frac{p}{k_B T}, \quad c_n = -\frac{U_n}{Vk_B T} + \frac{1}{V}\ln\frac{\rho^{(n)}}{z^n}, \quad k_n = \frac{R_n(\alpha_n V)}{R_{n-1}(\alpha_n V)} \cong 1, \quad \alpha_n = \alpha_0 + c_n, \quad c_n \cong (y_{f1}...y_{fn} - 1)\rho(z). \quad (29)$$

Подстановка выражения (29) в (28) приводит к трансцендентному уравнению для $\alpha_0$ вида

$$\frac{\alpha_0 k_n}{\alpha_0 + c_n} = -\frac{1}{8}\ln(1 - 8\rho \frac{k_n}{\alpha_0 + c_n}) - \frac{\alpha \rho^2}{k_B T}\frac{k_n}{\alpha_0 + c_n}. \qquad (30)$$

Для частиц с твердой сердцевиной из (25), (28) и (29) получаем

$$\alpha_0(d + \frac{k_{nd}}{(\alpha_0 + c_n)}) = -\frac{1}{8}\ln(1 - 8\rho(d + \frac{k_{nd}}{(\alpha_0 + c_n)})) - \frac{\alpha \rho^2}{k_B T}(d + \frac{k_{nd}}{(\alpha_0 + c_n)}), \quad k_{nd} = k_n(V \to V - nd) \quad . \quad (31)$$

При малых значениях $\rho$, $\alpha$, получаем из (30) и (31) уравнение состояния идеального газа $\alpha_0 = \rho$.

Величины $c_n$ и $\alpha_0 = p(z)/kT$ зависят от координат $r_1,...,r_n$ всех взаимодействующих частиц. От этой зависимости можно избавиться, предположив, что в равновесии система однородная, плотности $\rho^*(zy_{f1}...y_{fn})$ и $\rho$ не зависят от координат $r_1,...,r_n$, проинтегрировать выражения вида (24), (29) по $r_1,...,r_n$ и заменить $c_y$ на $c_{\bar{y}} \cong \int_V ... \int_V (y_{f1}...y_{fn} - 1) dr_1...dr_n / V^n$.

В первом приближении в [9] предполагается, что фактор исключения линейно уменьшается с увеличением плотности

$$f(\varphi) = f_0 - k_1 \rho v_0, \quad k_1 = (f_0/2)^2, \quad \varphi = \rho v_0, \quad f_0 = 8 \ . \tag{32}$$

Тогда в [9] из производящего уравнении получено уравнение Ван-дер-Ваальса, которое можно переписать в виде

$$\alpha_0 = \frac{\rho}{1 - 4\rho v_0} - \frac{\alpha \rho^2}{k_B T} \ . \tag{33}$$

Подставляя сюда значение $v_0$ из (29), получим квадратное уравнение для $\alpha_0$ с решением

$$\alpha_0 = \frac{1}{2} D_1 \pm \sqrt{(\frac{1}{2} D_1)^2 + \rho c_n + \frac{\alpha \rho^2}{k_B T}(4\rho k_n - c_n)}, \quad D_1 = \rho - \alpha \rho^2 / k_B T + 4\rho k_n - c_n \ . \tag{34}$$

В (34) входит параметр $k_n$ (15), куда включена и величина $\alpha_0$. Для систем большого объема $k_n \to 1$. Для частиц с твердой сердцевиной объемом $d$

$$\alpha_0 = \frac{1}{2} D_d \pm \sqrt{(\frac{1}{2} D_d)^2 + \frac{\rho [c_n + 4k_{nd}(\alpha \rho^2 / kT)]}{1 - 4\rho d} - c_n \frac{\alpha \rho^2}{kT}}, \quad D_d = \frac{\rho(1 + 4k_{nd})}{(1 - 4\rho d)} - \frac{\alpha \rho^2}{kT} - c_n \ . \tag{35}$$

При выборе знака плюс в (34), (35), как и в (30), получаем из (34) уравнение состояния идеального газа $\alpha_0 = \rho$. Из уравнений (34)-(35) можно получить критические параметры. Они выражаются через параметр взаимодействия $\alpha$, как в уравнении Ван-дер-Ваальса [3], и через параметр $c_y$ (24), через который выражен средний размер частиц (15), а также через объем твердой сердцевины $d$ в (35).

Во втором приближении в выражении для зависимости фактора исключения $f(\varphi)$ от общей объемной доли всех компонентов $\varphi$ в [9] моделируется вогнутость кривой зависимости $f(\varphi)$ с использованием дробно-линейной функции вида

$$f(\varphi) = \frac{f_0 - k_2 \varphi}{1 + k_3 \varphi}, \ k_2 + k_3 f_0 = K = 34, \ k_2 = (f_0 - K^{1/2}), \ k_3 = 2K^{1/2} - f_0, \ f_0 = 8, \beta_1 = f_0 - K^{1/2} \approx 2.169. \tag{36}$$

Уравнение состояния во втором приближении для неионной системы с ван-дер-ваальсовскими силами для однокомпонентного случая (значения параметров приведены в (36)) принимает вид

$$\alpha_0 = \frac{2.688\rho}{1-2.169\rho v_0} + \frac{0.778}{v_0}\ln(1-2.169\rho v_0) - \frac{\alpha\rho^2}{k_B T}. \tag{37}$$

Подстановка в (37) значения $v_0$ из (29) приводит, как в случае (30), к трансцендентному уравнению для $\alpha_0$ вида

$$\alpha_0 = \frac{2.688\rho(\alpha_0+c_n)}{\alpha_0+c_n-2.169\rho k_n} + \frac{0.778(\alpha_0+c_n)}{k_n}\ln(1-2.169\rho\frac{k_n}{\alpha_0+c_n}) - \frac{\alpha\rho^2}{k_B T}, \tag{38}$$

которое также переходит в уравнение состояния идеального газа при стремлении к нулю плотности числа частиц. Уравнение, аналогичное уравнению (31), записывается и для систем с твердой сердцевиной.

Дальнейшие приближения выше второго в [9] уже не содержат логарифмов и выражены общей формулой

$$\alpha_0^{(k)}v_0 = \frac{1}{(1-k_4\varphi)^{k-1}}\{\varphi + \frac{f_0-2k_4(k-1)}{(k-1)(k-2)k_4^2}[(1-k_4\varphi)^{k-1}-1+(k-1)k_4\varphi]\} - \frac{\alpha\rho^2 v_0}{k_B T}, \quad k\geq 3. \tag{39}$$

При $k=3$ уравнение (39) принимает вид

$$\alpha_0 v_0 = \frac{1}{(1-k_4\rho v_0)^2}\{\rho v_0 + \frac{f_0-4k_4}{2k_4^2}[(1-k_4\rho v_0)^2-1+2k_4\rho v_0]\} - \frac{\alpha\rho^2 v_0}{k_B T}, \tag{40}$$

после подстановки (29) в (40) получаем уравнение третьей степени для $\alpha_0$ вида

$$\alpha_0^3 + \alpha_0^2[2(c_n-k_4\rho k_n)-\rho+\alpha\rho^2/k_B T] + \tag{41}$$

$$+\alpha_0[(c_n-k_4\rho k_n)^2 - 2(\alpha\rho^2/k_B T)(c_n-k_4\rho k_n) - \rho(2c_n+k_n k_4^2 k`\rho)] +$$

$$+(\alpha\rho^2/k_B T)(c_n-k_4\rho k_n)^2 - c_n\rho(c_n+k`k_4^2\rho k_n) = 0, \quad k` = 2\frac{2-k_4}{k_4^2}.$$

В [8] уравнение состояния записывается в терминах фактора сжимаемости

$$Z = \frac{p}{\rho k_B T} = \frac{\alpha_0}{\rho}. \tag{42}$$

Для флюида твердых шаров одного размера, когда $f_0 = 8$, $\tilde{p}_1 = \alpha\rho^2 v_0/k_B T = 0$, выражение (39) представляется в виде

$$Z^{(k)} = \frac{1}{(1-k_4\varphi)^{k-1}}\{1 + \frac{8-2k_4(k-1)}{(k-1)(k-2)k_4^2\varphi}[(1-k_4\varphi)^{k-1}-1+(k-1)k_4\varphi]\}, \quad k\geq 3. \tag{43}$$

Отсюда можно получить уравнение Карнахэна-Старлинга

$$Z = \frac{1+\varphi+\varphi^2-\varphi^3}{(1-\varphi)^3}, \tag{44}$$

уравнения Перкуса-Йевика

$$Z = \frac{1+\varphi+\varphi^2}{(1-\varphi)^3}, \qquad Z = \frac{1+2\varphi+3\varphi^2}{(1-\varphi)^2} \tag{45}$$

(в вариантах уравнений сжимаемости и давления соответственно), и уравнение Гуггенгейма

$$Z = (1-\varphi)^{-4}. \tag{46}$$

Для четвертого и пятого приближений получаем

$$Z^{(4)} = \frac{1+(4-3k_4)\varphi + k_4(k_4 - 4/3)\varphi^2}{(1-k_4\varphi)^3}, \tag{47}$$

$$Z^{(5)} = \frac{1+(2/3)(1-k_4)(6\varphi - 4k_4\varphi^2 + k_4^2\varphi^3)}{(1-k_4\varphi)^4}. \tag{48}$$

При $k_4 = 1$ уравнение (47) близко к уравнению Карнахэна-Старлинга (44) и уравнению Перкуса-Йевика, а (48) совпадает с уравнением Гуггенгейма (46). Подстановка $v_0$ в виде (29) в (47) и (48) приводит к алгебраическим уравнениям для $\alpha_0$ четвертого и пятого степеней соответственно. В [9] постоянная $k_4$ используется как параметр подгонки к результатам численного эксперимента для всех плотностей. Уже в третьем приближении получен охват всех возможных плотностей. В [9] отмечено шестое приближение, которое при $k_4 \approx 0.803$ можно приближенно записать в виде

$$Z = (1-k_4\varphi)^{-5}. \tag{49}$$

Это простое уравнение оказывается наиболее точным из всех приближений (с максимальным отклонением от баз данных [9] не более 1%). Подстановка выражения (29) в (49) приводит к алгебраическому уравнению шестой степени для $\alpha_0$ вида

$$\alpha_0(\alpha_0 + c_n - \rho k_4 k_n)^5 = \rho(\alpha_0 + c_n)^5. \tag{50}$$

Для систем с твердой сердцевиной

$$\alpha_0[(\alpha_0 + c_n)(1 - k_4\rho d) - \rho k_4 k_n (V - nd)]^5 = \rho(\alpha_0 + c_n)^5.$$

## ЗАКЛЮЧЕНИЕ

Диаметр частиц $r_0$ занимает важное место в определении "размеров" сплошной среды [4]. Например, в приближении $\overline{\Delta^{(n)}}_1 \approx V/N$ трудно провести различие между микромасштабными и крупномасштабными флуктуациями [4]. В различных приложениях результаты настоящей работы можно использовать для систем, описывающихся распределением Гиббса. Так, в задачах физической химии или для коллоидных частиц – в той мере, в которой для них справедлива гиббсовская статистика. Полученные результаты могут оказаться полезными при исследовании ряда задач термодинамических свойств веществ. Основные результаты гиббсовской статистической физики (образование кластеров и различные приближения для уравнения состояния, отрицательность зависимости $\partial P/\partial v$, фазовые переходы, теория флуктуаций и т.д.) находят соответствие в определении эффективного объема частиц. Сопоставление результатов настоящей работы и работы [18] показывает их соответствие.

Размер частиц зависит от взаимодействий в системе. В настоящей работе рассмотрены гиббсовские распределения и равновесные состояния. Не оценивается возможное влияние неравновесных эффектов на размер частиц. Показательное распределение для размеров частиц получено из гиббсовской статистики. Если же исходить, например, из распределений Цаллиса [26], то распределение для размеров частиц будет иметь степенной характер.

В [9] размеры частиц учитываются при помощи задания различного вида фактора исключения. В настоящей работе проводится учет размеров частиц при помощи строгих соотношений статистической физики. Сопоставление с экспериментальными данными, например, уравнения Ван-дер-Ваальса приводит к выводу о зависимости от давления параметра, связанного с размером частиц. Такая зависимость учитывается в настоящей работе, при переходе от уравнения (33) к уравнениям (34), (35). И в других уравнениях состояния плохо определенная постоянная величина собственного объема частиц заменяется функцией давления, температуры и потенциала взаимодействия в системе. Выражение для $c_n$ (24) приближенное. Можно записать более точные выражения. Выражения $c_y$ и $c_{\bar{y}}$ из (24), (29) можно записать через вириальные коэффициенты. Точное выражение для $\alpha_n = \alpha_0 + c_n$ (18) неизвестно и зависит от активности. Подход этой работы, учет явных соотношений для размеров частиц, можно рассматривать как повышение точности описания. Само соотношение для среднего размера частиц (15) можно рассматривать как уравнение состояния.

Уравнения состояния для систем с твердой сердцевиной позволяют найти зависимости термодинамических параметров от величины твердой сердцевины $d$. Роль объема частиц в [9] и других работах, в которых рассматривается объем частиц, здесь играет величина $d$. В настоящей работе не получены новые уравнения состояния. Проводится уточнение известных уравнений состояния, своеобразная надстройка над ними.

# СПИСОК ЛИТЕРАТУРЫ

Рисунки к статье В. В. Рязанова «Уравнение состояния и распределение размеров частиц в гиббсовской системе».

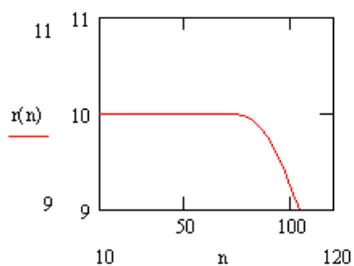

Рис. 1а. n=10,…,120; a=$10^{-1}$ ;  d=$10^{-5}$ ;  V=$10^{3}$ .

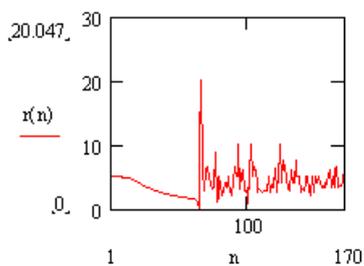

Рис.1б. n=1,2,…,170; a=$10^{-0.7}$ ;  d=$10^{-3}$;  V=$10^{2}$

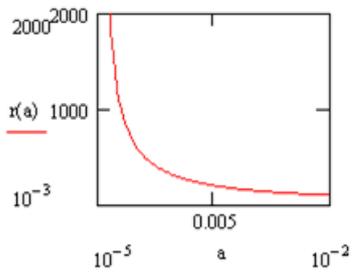

Рис.2а. $a=10^{-5}, 10^{-4}, …, 10^{-2}$ ; $d=10^{-7}$;  $n=10$ ; $V=10^{17}$

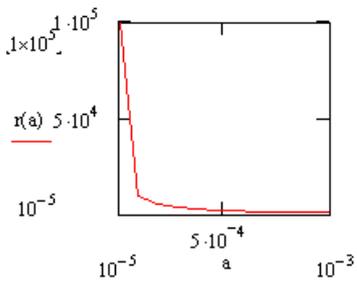

Рис. 2б.  $a=10^{-5}, 10^{-4}; …,10^{-3}$;   $d=10^{-5}$;  $V=10^{10}$; $n=30$ .

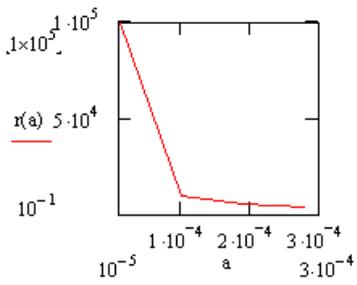

Рис. 2в.  $a=10^{-5},…,3·10^{-4}$;   $d=10^{-1}$;  $V=10^{10}$ ; $n=47$ .

Подписи к рисункам к статье В. В. Рязанова «Уравнение состояния и распределение размеров частиц в гиббсовской системе».

Рис.1. Зависимость разности между средним размером частицы $\overline{\Delta^{(n)}{}_{1\,d}}$ в кластере n взаимодействующих частиц в модели с жесткой несжимаемой сердцевиной объемом d и величиной d от n - числа частиц в группе взаимодействующих частиц. Использована модель взаимодействия упругих шаров [2]; $\partial \overline{\Delta^{(n)}{}_{1\,d}}/\partial n \leq 0$; $r = \overline{\Delta^{(n)}{}_{1\,d}} - d$; $a = \alpha_n = P(zy_1...y_n)/k_B T$.

Рис. 1а: $\partial \overline{\Delta^{(n)}{}_{1\,d}}/\partial n \leq 0$; $a=10^{-1}$; $d=10^{-5}$; $V=10^3$; $10 \leq n \leq 120$;     Рис. 1б: при некоторых n $\partial \overline{\Delta^{(n)}{}_{1\,d}}/\partial n > 0$, область фазового перехода; $a=10^{-0.7}$; $d=10^{-3}$; $V=10^2$; $1 \leq n \leq 170$.

Рис.2. Зависимость разности между средним размером частицы $\overline{\Delta^{(n)}{}_{1\,d}}$ в кластере n взаимодействующих частиц в модели с жесткой несжимаемой сердцевиной объемом d и величиной d от $a = \alpha_n$ – фактора сжимаемости (умноженного на $R/k_B V$, R – универсальная газовая постоянная); $r = \overline{\Delta^{(n)}{}_{1\,d}} - d$; $a = \alpha_n = P(zy_1...y_n)/k_B T$.

Рис. 2а: $10^{-5} \leq a \leq 10^{-2}$; $d=10^{-7}$; $n=10$; $V=10^{17}$. Рис. 2б: $10^{-5} \leq a \leq 10^{-3}$; $d=10^{-5}$; $n=30$; $V=10^{10}$. Рис. 2в: $10^{-5} \leq a \leq 3 \cdot 10^{-4}$; $d=10^{-1}$; $n=47$; $V=10^{10}$.